\documentclass{ws-ijmpd}

\begin{document}

\markboth{Bamba, Geng and Lee}
{Phantom crossing in viable $f(R)$ theories}

%
\catchline{}{}{}{}{}
%

\title{Phantom crossing in viable $f(R)$ theories}

\author{Kazuharu Bamba$^a$\footnote{Present address:Kobayashi-Maskawa Institute for the Origin of Particles and the Universe, 
Nagoya University, Nagoya 464-8602, Japan.}, Chao-Qiang Geng$^{a,b}$
 and Chung-Chi Lee$^a$\footnote{Talk presented by C.C. Lee at the
 2nd International Workshop on Dark Matter, Dark Energy and Matter-Antimatter Asymmetry, Hsinchu, Taiwan, 5-6 Nov 2010.}}

\address{$^a$Department of Physics, National Tsing Hua University,
 Hsinchu, Taiwan, R.O.C.
\\
$^b$National Center for Theoretical Sciences,
 Hsinchu, Taiwan, R.O.C.\\
E-Mail addresses: bamba@kmi.nagoya-u.ac.jp, geng@phys.nthu.edu.tw, g9522545@oz.nthu.edu.tw}

\maketitle

\begin{abstract}

We review the equation of state for dark energy 
in modified gravity theories. In particular, we summarize 
the generic feature of the phantom divide crossing in the past and future in viable $f(R)$ gravity models.

\end{abstract}



To understand the late time acceleration universe~\cite{Copeland:2006wr}, one of the interesting possibilities 
 is to consider a modified gravitational theory, 
such as $f(R)$ gravity~\cite{Sotiriou:2008rp}. To build up a
  viable $f(R)$ gravity model, one needs to 
satisfy the following conditions: 
(a) positivity of the effective gravitational coupling, 
(b) stability of cosmological perturbations~\cite{Nojiri:2003ft}, 
(c) asymptotic behavior to the standard $\Lambda$-Cold-Dark-Matter 
($\Lambda\mathrm{CDM}$) model in the large curvature regime, 
(d) stability of the late-time de Sitter point~\cite{Muller:1987hp}, 
(e) constraints from the equivalence principle, 
and 
(f) solar-system constraints~\cite{Solar-System-Constraints}. 
%
Several viable models 
have been constructed in the literature, 
such as the following popular ones:\cite{Hu:2007nk,Starobinsky:2007hu,Tsujikawa:2007xu,Exponential-type-f(R)-gravity,Cognola:2007zu,Linder:2009jz,Bamba:2010ws,BGL} 
{\begin{tabular}{@{}ccccc@{}}
\\
&& model & $f(R)$ & Constant parameters \\
\\
&&(i) Hu-Sawicki & $ R - \frac{c_1 R_{\mathrm{HS}} \left(R/R_{\mathrm{HS}}\right)^p}{c_2 
\left(R/R_{\mathrm{HS}}\right)^p + 1}$ &$c_1$, $c_2$, $p(>0)$,$R_{\mathrm{HS}}(>0)$\\
&&(ii) Starobinsky& $ R + \lambda R_{\mathrm{S}} \left[\left(1+\frac{R^2}{R_{\mathrm{S}}^2} \right)^{-n}-1 \right]$ & $\lambda (>0)$, $n (>0)$, 
$R_{\mathrm{S}}$\\
&&(iii) Tsujikawa&$R - \mu R_{\mathrm{T}} \tanh\left( \frac{R}{R_{\mathrm{T}}} \right)$
&$\mu (>0)$, $R_{\mathrm{T}} (>0)$\\
&&(iv) Exponential & $R -\beta R_{\mathrm{E}}\left(1-e^{-R/R_{\mathrm{E}}} 
\right)$ & $\beta$, $R_{\mathrm{E}}$
\\
\\
\end{tabular} \label{Table}}

Recently, 
the cosmological observational data~\cite{observational status}  seems to 
indicate the crossing of the phantom divide $w_{\mathrm{DE}}=-1$ of the 
equation of state for dark energy 
in the near past. To understand such a crossing, many attempts have been 
made. 
The most noticeable one is to use a phantom field with a negative kinetic 
energy term~\cite{Phantom}. 
Clearly, it surfers a serious problem as it is not stable at the quantum 
level. 
On the other hand, 
the crossing of the phantom divide can also be realized in 
the above viable $f(R)$ 
models~\cite{Hu:2007nk,Linder:2009jz,Bamba:2010ws,BGL,Martinelli:2009ek} 
without violating any stability conditions.
This is probably 
the most peculiar character of the modified gravitational models. 
Other $f(R)$ gravity models  with realizing 
a crossing~\cite{Abdalla:2004sw} 
as well as multiple crossings~\cite{Bamba:2009kc} of 
the phantom boundary have also been examined.

In this talk, 
we would like to review equation of state in $f(R)$ gravity. 
In particular, 
we show that the viable $f(R)$ models 
generally exhibit 
the crossings of the phantom divide in the past as well as future~\cite{Bamba:2010ws,BGL}.

The action of $f(R)$ gravity with matter is given by
\begin{equation}
I = \int d^4 x \sqrt{-g} \frac{f(R)}{2\kappa^2} + I_{\mathrm{matter}} 
(g_{\mu\nu}, \Upsilon_{\mathrm{matter}})\,,
\label{eq:1}
\end{equation} 
where $g$ is the determinant of the metric tensor $g_{\mu\nu}$, 
$I_{\mathrm{matter}}$ is the action of matter which is assumed to be minimally 
coupled to gravity, i.e., the action $I$ is written in the Jordan frame, 
and $\Upsilon_{\mathrm{matter}}$ denotes matter fields. 
Here, we use the standard metric formalism. 
By taking the variation of the action in Eq.~(\ref{eq:1}) with respect to 
$g_{\mu\nu}$, one obtains~\cite{Sotiriou:2008rp} 
\begin{equation}
F G_{\mu\nu} 
= 
\kappa^2 T^{(\mathrm{matter})}_{\mu \nu} 
-\frac{1}{2} g_{\mu \nu} \left( FR - f \right)
+ \nabla_{\mu}\nabla_{\nu}F -g_{\mu \nu} \Box F\,,
\label{eq:2}
\end{equation}
where 
$G_{\mu\nu}=R_{\mu\nu}-\left(1/2\right)g_{\mu\nu}R$ is 
the Einstein tensor, 
$F(R) \equiv d f(R)/dR$, 
${\nabla}_{\mu}$ is the covariant derivative operator associated with 
$g_{\mu \nu}$, 
$\Box \equiv g^{\mu \nu} {\nabla}_{\mu} {\nabla}_{\nu}$
is the covariant d'Alembertian for a scalar field, 
and 
$T^{(\mathrm{matter})}_{\mu \nu}$ 
is the contribution to the energy-momentum tensor from all 
perfect fluids of matter. 

{}From Eq.~(\ref{eq:2}), 
we obtain the following gravitational field equations: 
\begin{eqnarray}
3FH^2 
&=&
\kappa^2 \rho_{\mathrm{M}} +\frac{1}{2} \left( FR - f \right) 
-3H\dot{F}\,,
 \nonumber\\ 
-2F\dot{H}  
&=&
\kappa^2 \left( \rho_{\mathrm{M}} + P_{\mathrm{M}} \right)
+\ddot{F}-H\dot{F}\,,
\label{eq:3-3}
\end{eqnarray}
where $H=\dot{a}/a$ is the Hubble parameter, 
the dot denotes the time derivative of $\partial/\partial t$, and 
$\rho_{\mathrm{M}}$ and $P_{\mathrm{M}}$ are 
the energy density and pressure of all perfect fluids of matter, 
respectively. 

The equation of state for dark energy is given by
\begin{equation} 
w_{\mathrm{DE}} \equiv P_{\mathrm{DE}}/\rho_{\mathrm{DE}},
\end{equation}
where
\begin{eqnarray}
 \rho_{\mathrm{DE}}& =& \frac{1}{\kappa^2} \left[ 
\frac{1}{2} \left( FR - f \right) 
-3H \dot{F} 
+3\left(1-F\right)H^2 
\right]\,,
\nonumber\\
 P_{\mathrm{DE}} &=& \frac{1}{\kappa^2} 
\left[ -\frac{1}{2} \left( FR - f \right) 
+\ddot{F}+2H \dot{F}
-\left(1-F\right)\left(2\dot{H}+3H^2\right) 
\right]\,.
 \label{eq:W}
\end{eqnarray}\\
\begin{center}
\begin{figure}[pb]
\begin{tabular}{ll}
\begin{minipage}{60mm}
\begin{center}
\unitlength=1mm
\resizebox{!}{5.5cm}{
   \includegraphics{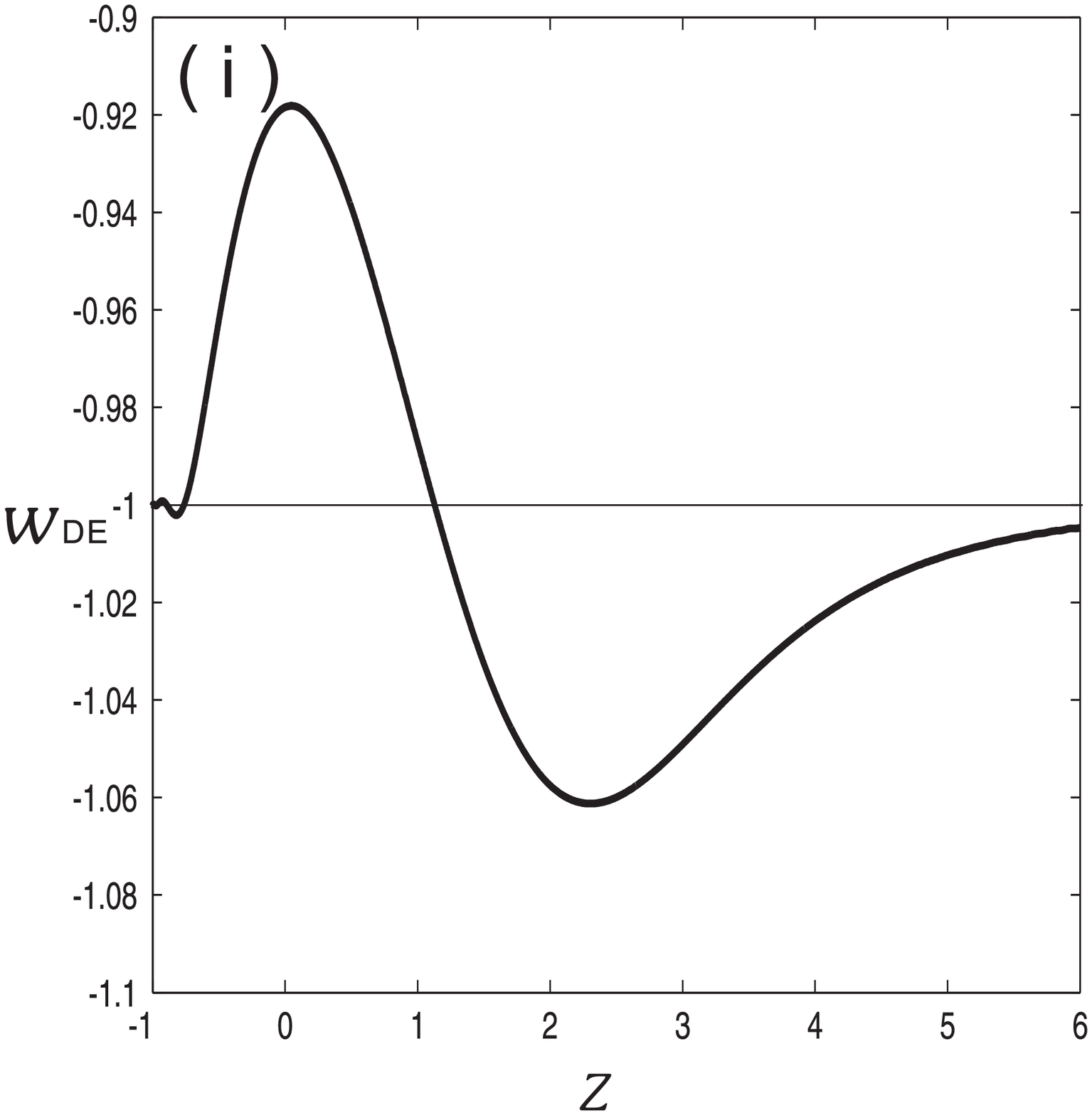}
                  }
\end{center}
\end{minipage}

\begin{minipage}{60mm}
\begin{center}
\unitlength=1mm
\resizebox{!}{5.5cm}{
   \includegraphics{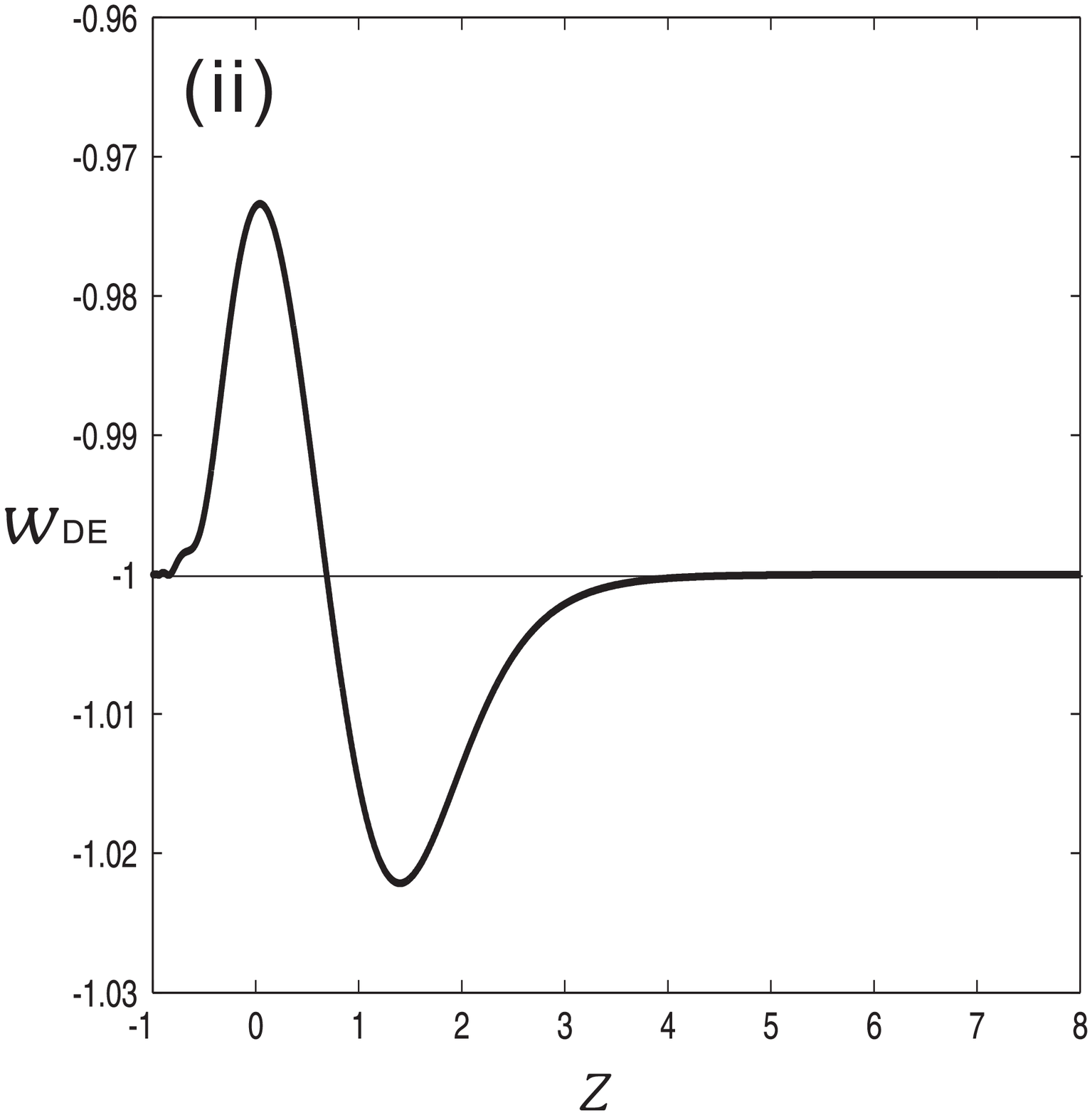}
                  }
\end{center}
\end{minipage}\\[1mm]

\begin{minipage}{60mm}
\begin{center}
\unitlength=1mm
\resizebox{!}{5.5cm}{
   \includegraphics{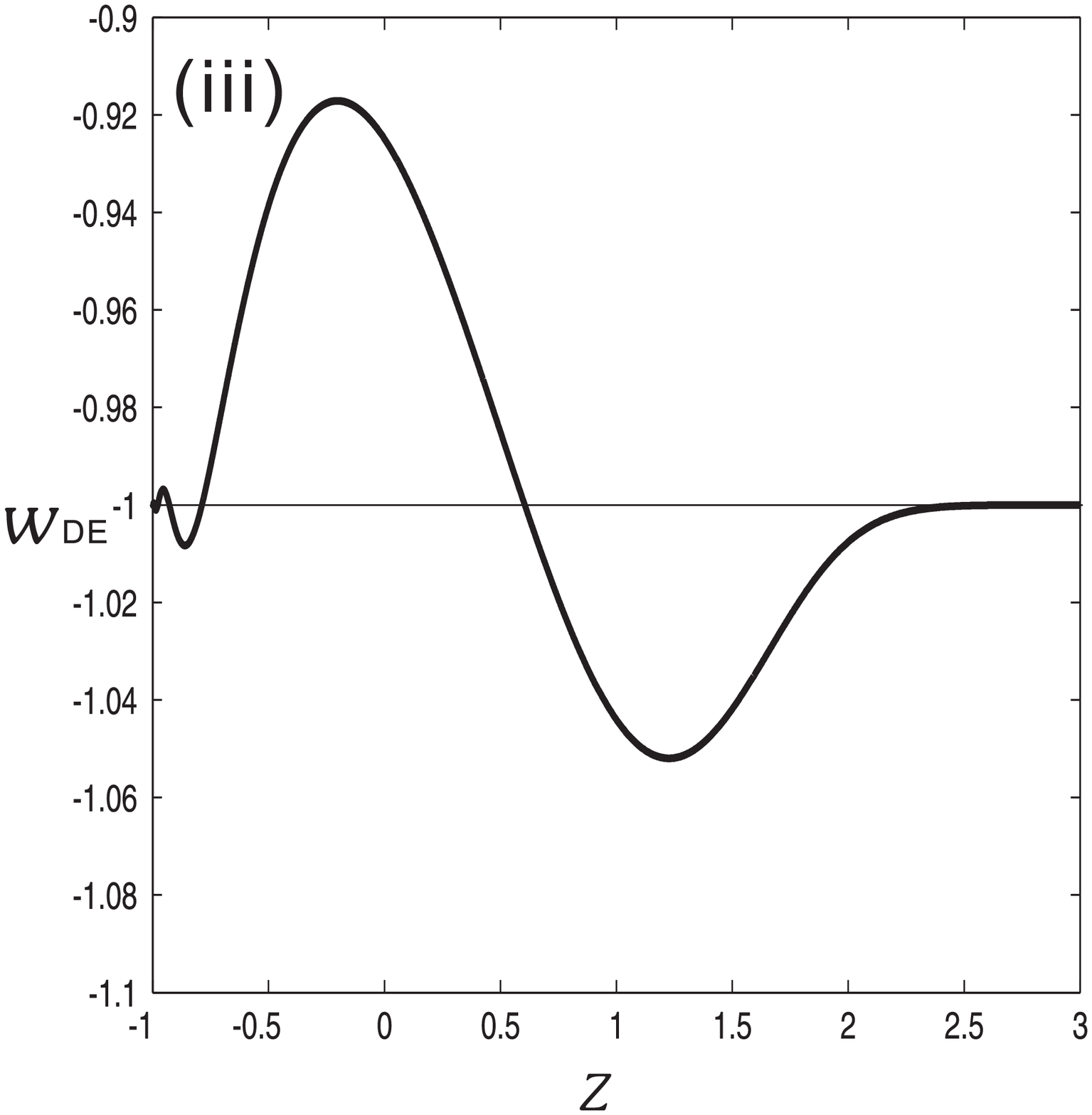}
                  }
\end{center}
\end{minipage}

\begin{minipage}{60mm}
\begin{center}
\unitlength=1mm
\resizebox{!}{5.5cm}{
   \includegraphics{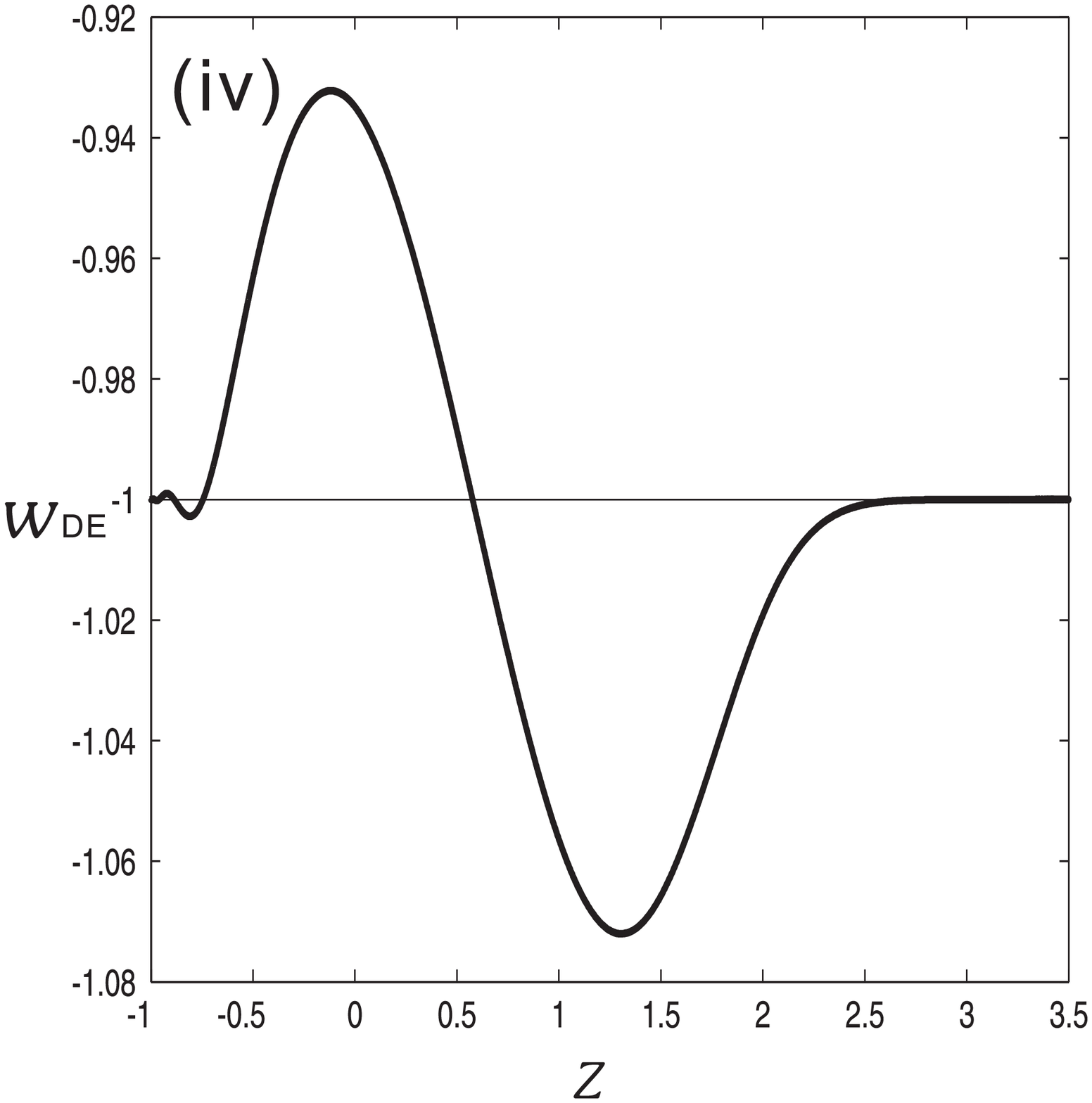}
                  }
\end{center}
\end{minipage}
\end{tabular}
\caption{Evolutions of the equation of state
$w_{\mathrm{DE}}$ 
as functions of the redshift $z$ 
in 
(i) Hu-Sawicki model for $p=1$, $c_1=2$ and $c_2=1$, 
(ii) Starobinsky model for $n=2$ and $\lambda=1.5$, 
(iii) Tsujikawa model for $\mu=1$
and 
(iv) the exponential gravity model for $\beta=1.8$, 
respectively, where
the thin solid lines show $w_{\mathrm{DE}}=-1$ (cosmological constant). 
}
\label{fg:1}
\end{figure}
\end{center}

\begin{center}
\begin{figure}[tbp]
\begin{tabular}{ll}
\begin{minipage}{60mm}
\begin{center}
\unitlength=1mm
\resizebox{!}{5.5cm}{
   \includegraphics{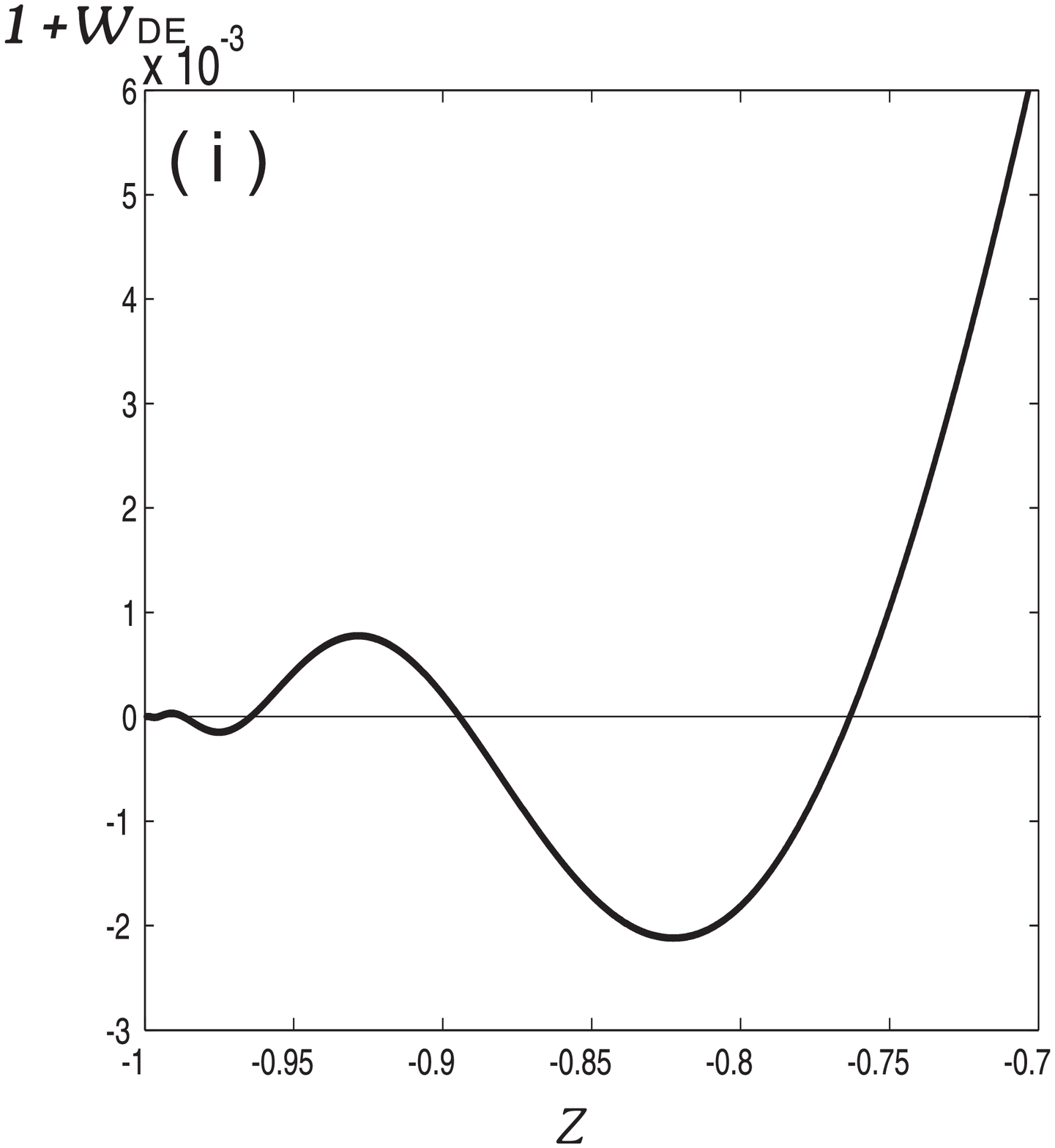}
                  }
\end{center}
\end{minipage}

\begin{minipage}{60mm}
\begin{center}
\unitlength=1mm
\resizebox{!}{5.5cm}{
   \includegraphics{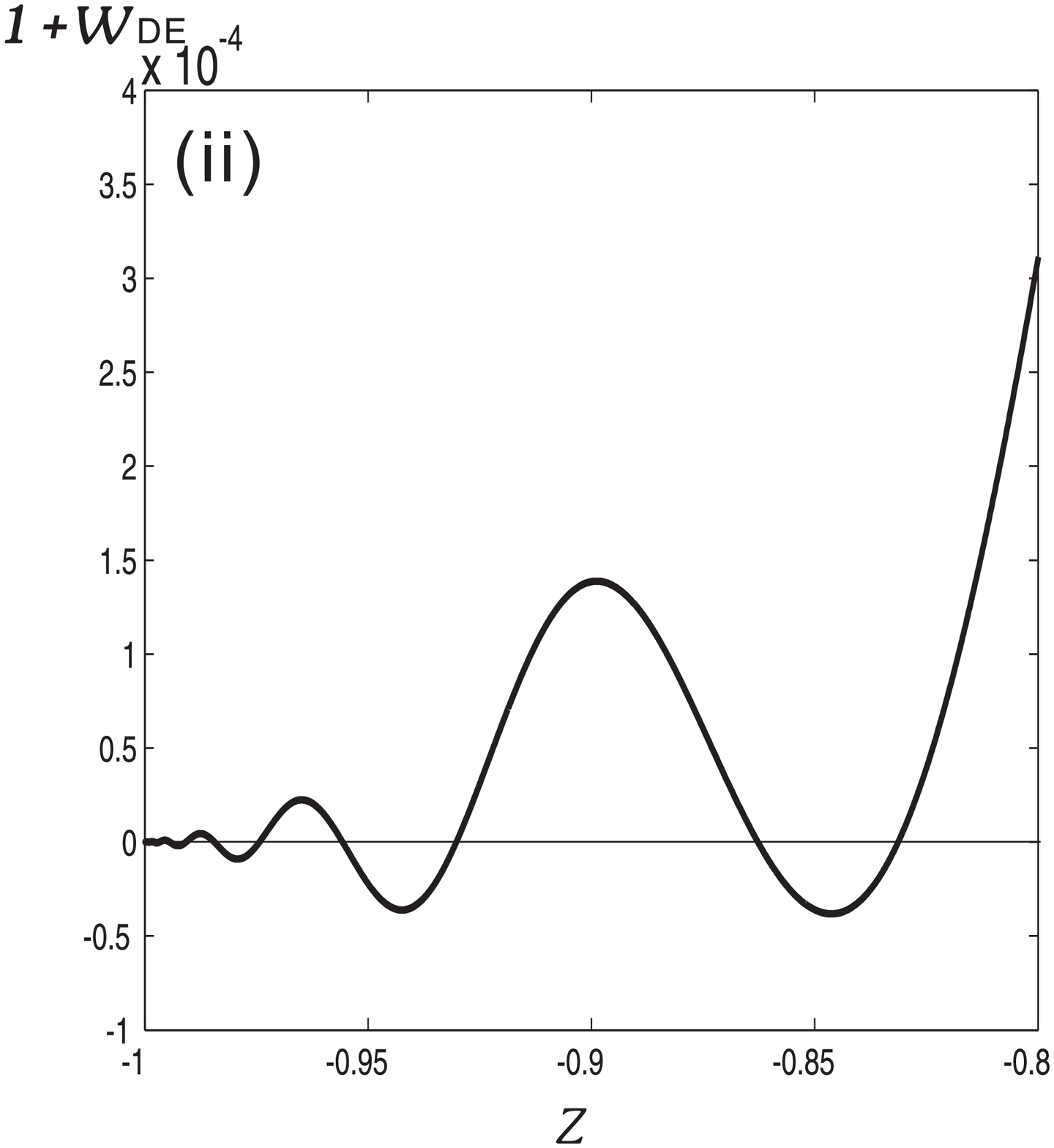}
                  }
\end{center}
\end{minipage}\\[1mm]

\begin{minipage}{60mm}
\begin{center}
\unitlength=1mm
\resizebox{!}{5.5cm}{
   \includegraphics{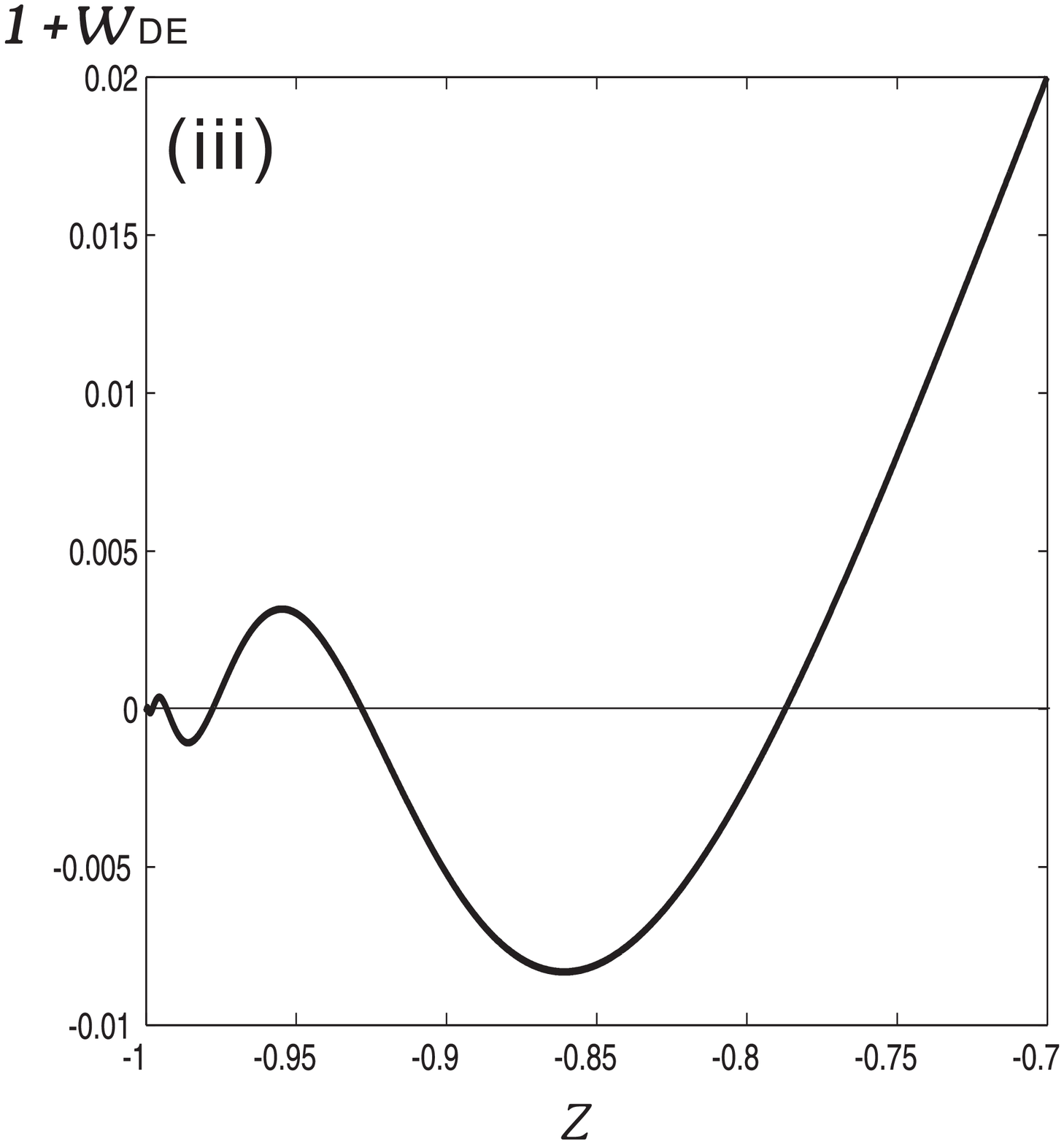}
                  }
\end{center}
\end{minipage}

\begin{minipage}{60mm}
\begin{center}
\unitlength=1mm
\resizebox{!}{5.5cm}{
   \includegraphics{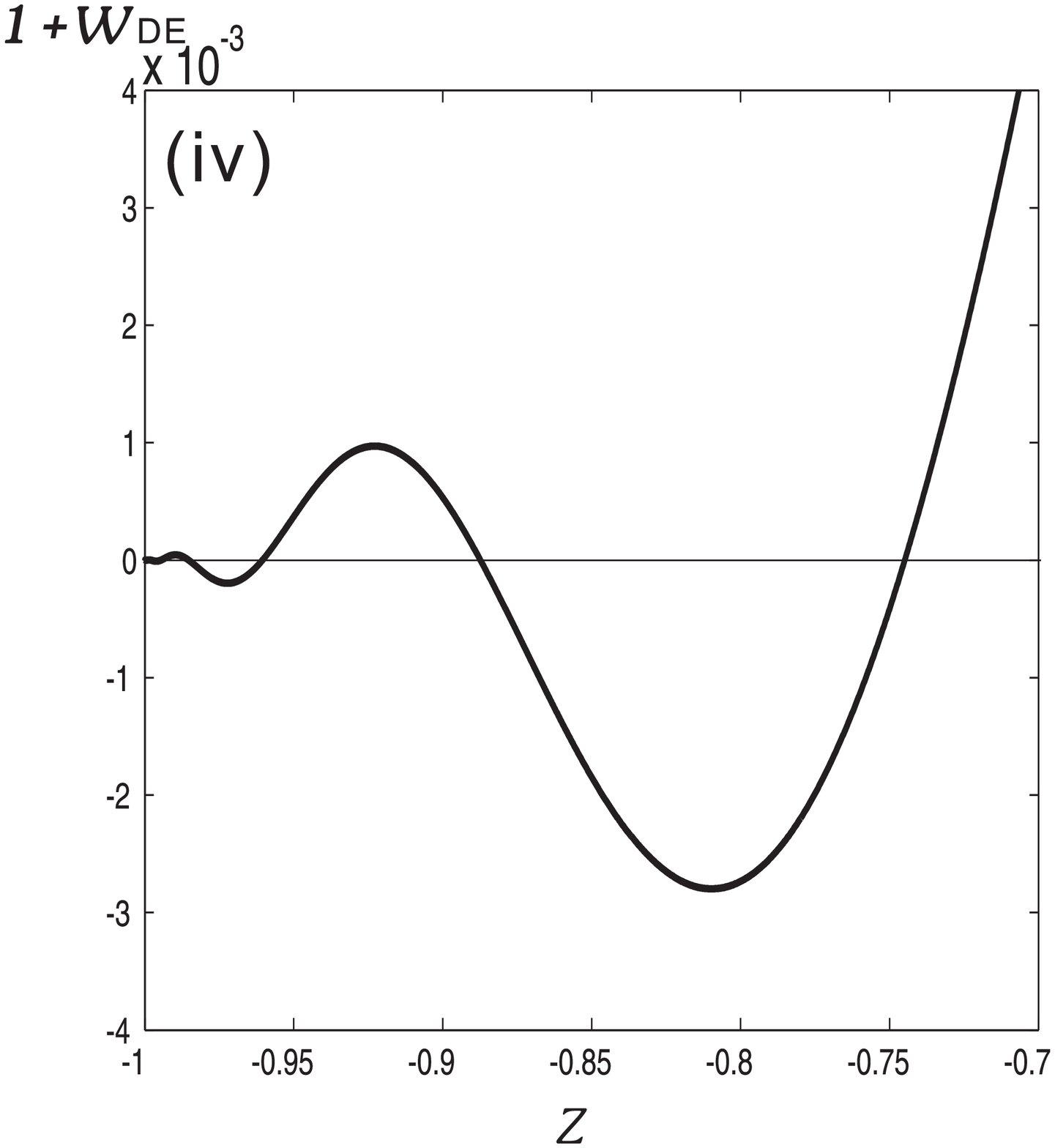}
                  }
\end{center}
\end{minipage}
\end{tabular}
\caption{Future evolutions of 
$1+w_{\mathrm{DE}}$ 
as functions of the redshift $z$. 
Legend is the same as Fig.~1.
}
\label{fg:2}
\end{figure}
\end{center}

\begin{center}
\begin{figure}[tbp]
\begin{tabular}{ll}
\begin{minipage}{60mm}
\begin{center}
\unitlength=1mm
\resizebox{!}{5.5cm}{
   \includegraphics{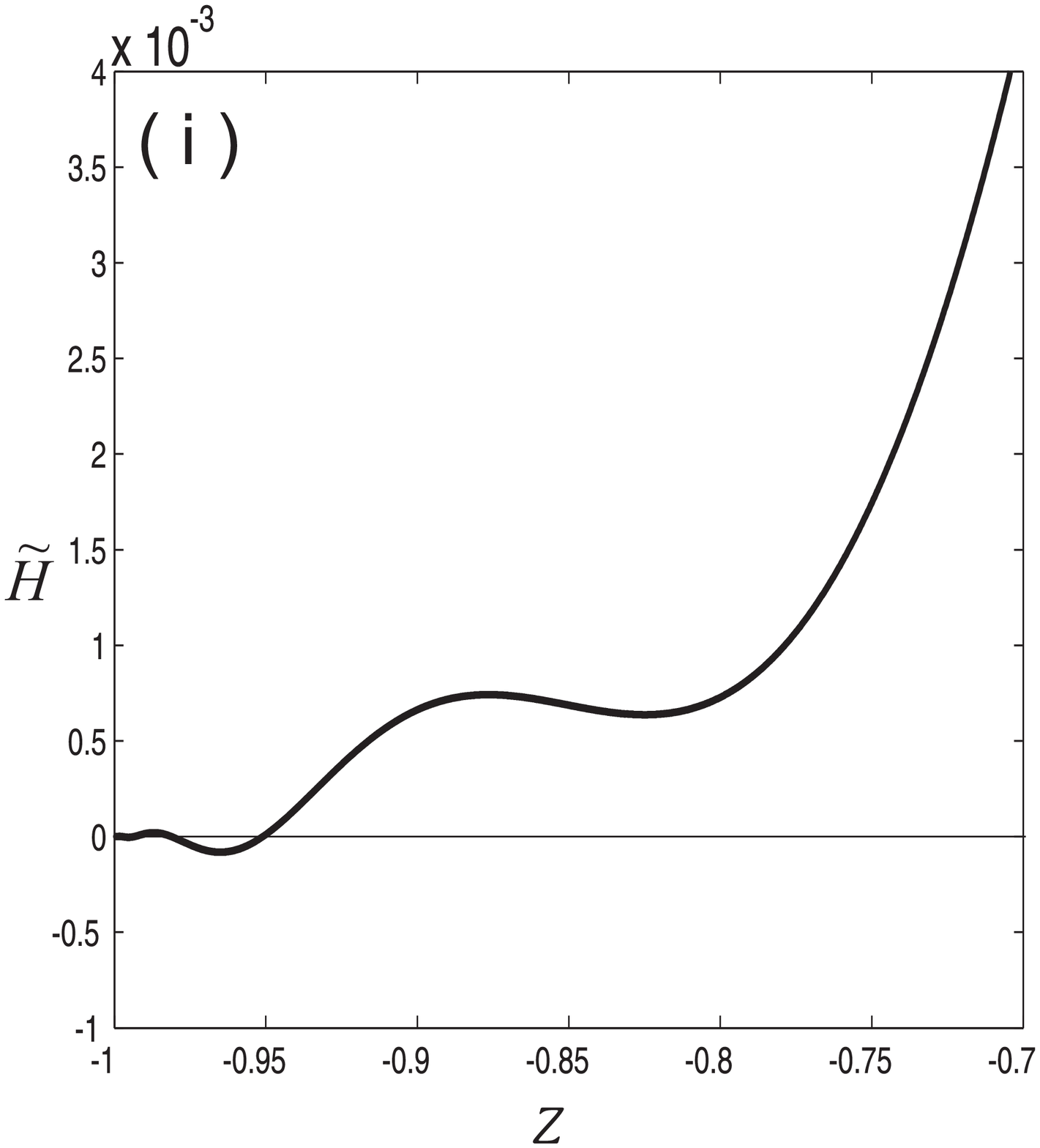}
                  }
\end{center}
\end{minipage}

\begin{minipage}{60mm}
\begin{center}
\unitlength=1mm
\resizebox{!}{5.5cm}{
   \includegraphics{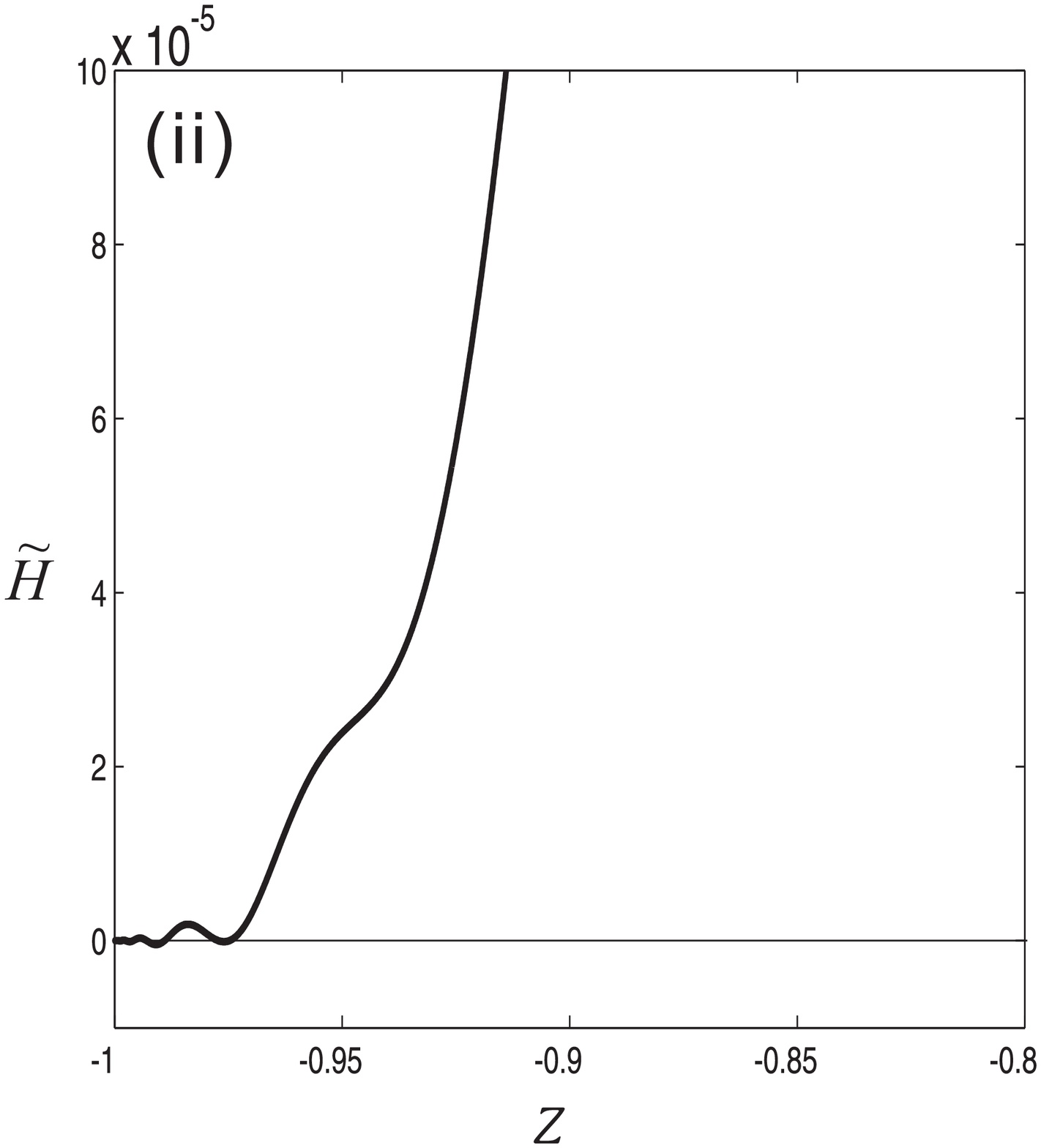}
                  }
\end{center}
\end{minipage}\\[1mm]

\begin{minipage}{60mm}
\begin{center}
\unitlength=1mm
\resizebox{!}{5.5cm}{
   \includegraphics{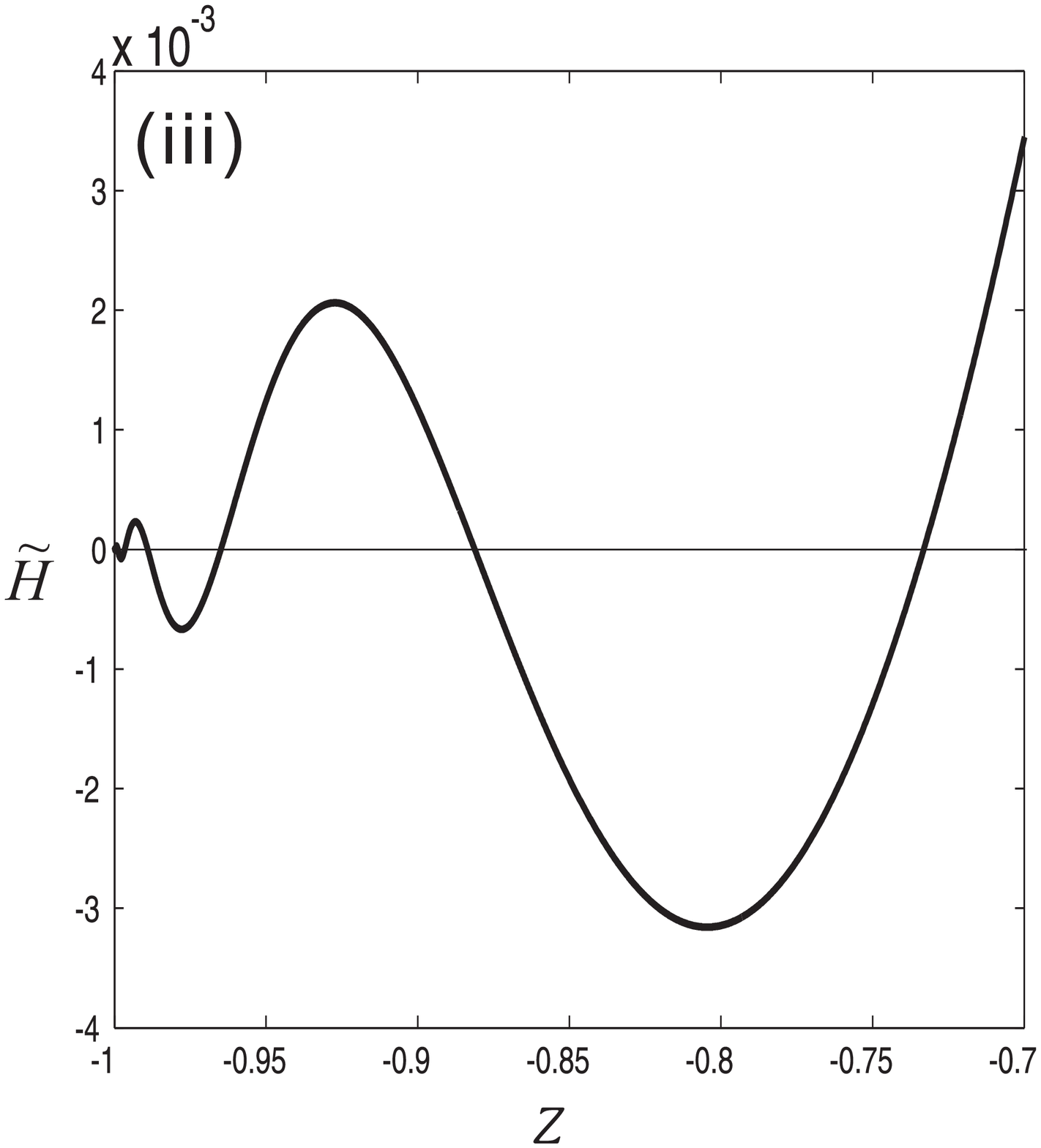}
                  }
\end{center}
\end{minipage}

\begin{minipage}{60mm}
\begin{center}
\unitlength=1mm
\resizebox{!}{5.5cm}{
   \includegraphics{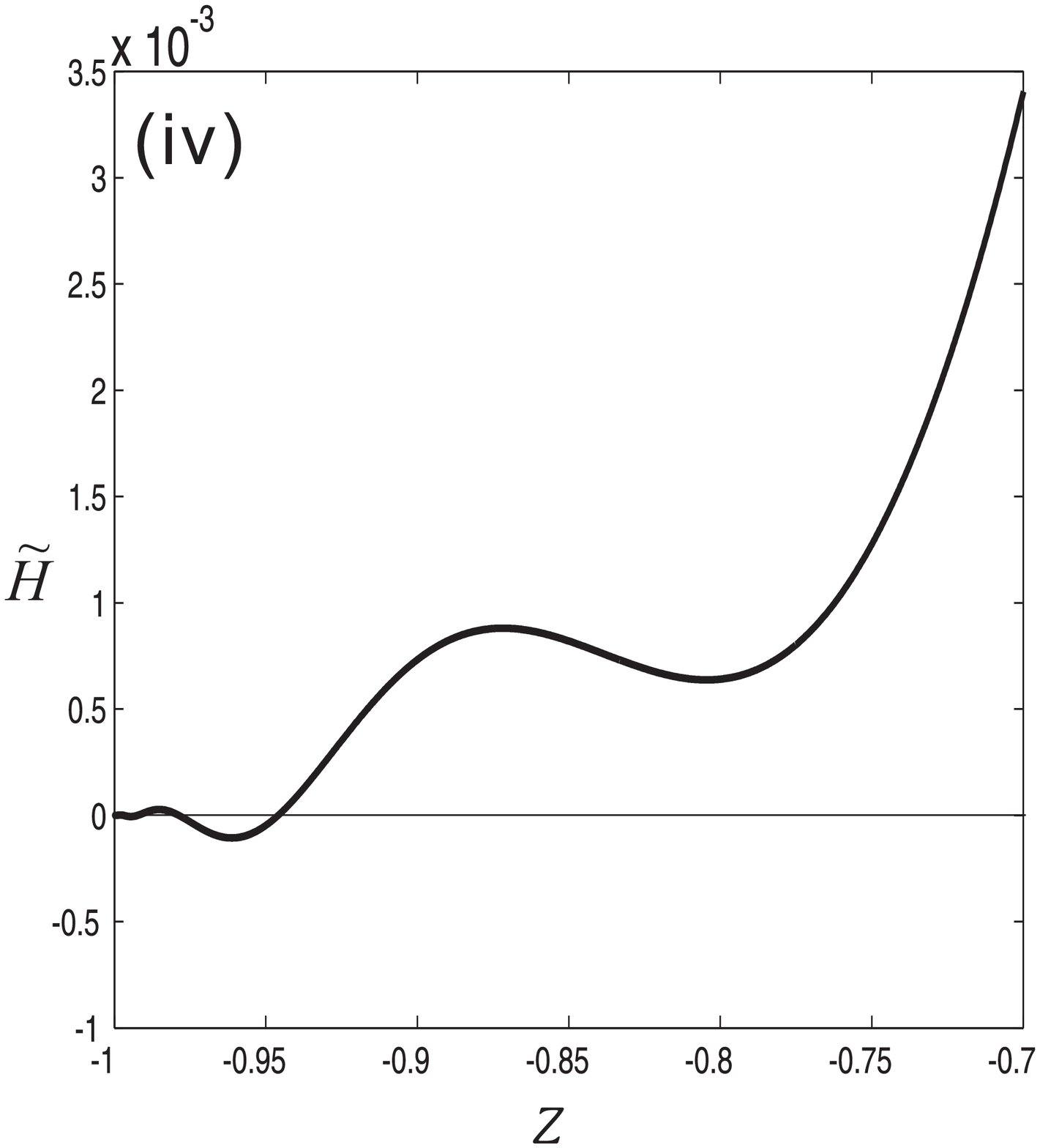}
                  }
\end{center}
\end{minipage}
\end{tabular}
\caption{Future evolutions of 
$\tilde{H} \equiv \bar{H} - \bar{H}_{\mathrm{f}}$ 
with $\bar{H} \equiv H/H_0$ and $\bar{H}_{\mathrm{f}} \equiv H(z=-1)/H_0$ 
as functions of the redshift $z$. 
Legend is the same as Fig.~1. 
}
\label{fg:3}
\end{figure}
\end{center}

\vspace{-18mm}

In Figs.~1, 2 and 3, 
we depict the evolution of $w_{\mathrm{DE}}$,
 future evolutions of $1+w_{\mathrm{DE}}$, 
and $\tilde{H} \equiv \bar{H} - \bar{H}_{\mathrm{f}}$ 
with $\bar{H} \equiv H/H_0$ and $\bar{H}_{\mathrm{f}} \equiv H(z=-1)/H_0$, 
as functions of the redshift $z \equiv 1/a -1$ 
in 
(i) Hu-Sawicki model for $p=1$, $c_1=2$ and $c_2=1$, 
(ii) Starobinsky model for $n=2$ and $\lambda=1.5$, 
(iii) Tsujikawa model for $\mu=1$
and 
(iv) the exponential gravity model for $\beta=1.8$, 
respectively, and the subscript `f' denotes the value at the final stage $z=-1$. 
Note that the present time is $z=0$ and the future is $-1 \leq z<0$. 
The parameters used for each model 
in Figs.~1--3 are the viable 
ones~\cite{Capozziello:2007eu,Tsujikawa:2009ku}. 
Several remarks are as follows:  
(a) the qualitative results do not strongly depend on the values of 
the parameters in each model; (b) we have studied the Appleby-Battye model~\cite{Appleby:2007vb}, which 
is also a viable $f(R)$ model, and we have found that the numerical results 
are similar to those in the Starobinsky model of (ii) as expected.

We note that the present values of  
$w_{\mathrm{DE}}(z=0)$ are  -0.92, -0.97, -0.92 and -0.93 
for the models of (i)--(iv), respectively. 
These values satisfy the present observational 
constraints~\cite{Komatsu:2010fb}. 
Moreover, 
a dimensionless 
quantity $H^2/\left(\kappa^2 \rho_{\mathrm{m}}^{(0)}/3\right)$ 
can be determined through the numerical calculations, 
where $\rho_{\mathrm{m}}^{(0)}$ is the energy density of 
non-relativistic matter at the present time. 
If we use the observational data on the current density parameter of 
non-relativistic matter 
$\Omega_{\mathrm{m}}^{(0)} \equiv 
\rho_{\mathrm{m}}^{(0)}/\rho_{\mathrm{crit}}^{(0)}
= 0.26$ with 
$\rho_{\mathrm{crit}}^{(0)} = 3H_0^2/\kappa^2$~\cite{Komatsu:2010fb}, 
we find that the present value of the Hubble parameter 
$H_0 = H(z=0)$ is $H_0 = 71 \mathrm{km/s/Mpc}$~\cite{Komatsu:2010fb} 
for all the models of (i)--(iv). 
Furthermore, 
$\bar{H}_{\mathrm{f}}$  = 0.80 , 0.85 , 0.78 and 0.81
for the models of (i)--(iv), respectively,
where $\bar{H}_{\mathrm{f}} \equiv H(z=-1)/H_0$. 

It is clear from Figs.~1--3 that 
in the future ($-1 \leq z \lesssim -0.74$), 
the crossings of the phantom divide 
are the generic feature for all the existing viable $f(R)$ models. 
By writing the first future crossing of the phantom divide 
and 
the first sign change of $\dot{H}$ from 
negative to positive as
$z = z_{\mathrm{cross}}$ 
and $z = z_{\mathrm{p}}$, respectively, we find that
 $(z_{\mathrm{cross}}, z_{\mathrm{p}})_{\alpha} = (-0.76, -0.82)_i$, 
$(-0.83, -0.98)_{ii}$, 
$(-0.79, -0.80)_{iii}$ and $(-0.74, -0.80)_{iv}$,
where the subscript $\alpha$ represents the $\alpha$th viable model. 
The values of the ratio 
$\Xi \equiv \Omega_{\mathrm{m}}/\Omega_{\mathrm{DE}}$ at 
$z = z_{\mathrm{cross}}$ and $z = z_{\mathrm{p}}$ are 
$(\Xi(z = z_{\mathrm{cross}}), \Xi(z = z_{\mathrm{p}}))_{\alpha} = 
(5.2 \times 10^{-3}, 2.1 \times 10^{-3})_i$, 
$(1.7 \times 10^{-3}, 4.8 \times 10^{-6})_{ii}$, 
$(4.1 \times 10^{-3}, 3.1 \times 10^{-3})_{iii}$ and 
$(6.2 \times 10^{-3}, 2.8 \times 10^{-3})_{iv}$, 
where 
$\Omega_{\mathrm{DE}} \equiv \rho_{\mathrm{DE}}/\rho_{\mathrm{crit}}^{(0)}$ 
and 
$\Omega_{\mathrm{m}} \equiv \rho_{\mathrm{m}}/\rho_{\mathrm{crit}}^{(0)}$ 
are the density parameters of dark energy and non-relativistic matter 
(cold dark matter and baryon), 
respectively. 
%
%
As $z$ decreases ($-1 \leq z \lesssim -0.90$), 
dark energy becomes much more 
dominant over non-relativistic matter 
($\Xi = \Omega_{\mathrm{m}}/\Omega_{\mathrm{DE}} \lesssim 10^{-5}$). 
As a result, one has $w_{\mathrm{DE}} \approx 
w_{\mathrm{eff}} \equiv 
-1 -2\dot{H}/\left(3H^2\right) 
= P_{\mathrm{tot}}/\rho_{\mathrm{tot}} 
$, 
where $w_{\mathrm{eff}}$ is the effective equation of state for the universe, 
and 
$\rho_{\mathrm{tot}} \equiv \rho_{\mathrm{DE}} + \rho_{\mathrm{m}} + 
\rho_{\mathrm{r}}$ and 
$P_{\mathrm{tot}} \equiv P_{\mathrm{DE}}  + P_{\mathrm{r}}$ 
are the total energy density and pressure of the universe, 
respectively. Here, $\rho_{\mathrm{m(r)}}$ 
and 
$P_{\mathrm{r}}$ are 
the energy density of non-relativistic matter (radiation) and 
the pressure of 
radiation, respectively. 
The physical reason why the crossing of the phantom divide appears 
in the farther future ($-1 \leq z \lesssim -0.90$) is that 
the sign of $\dot{H}$ 
changes from negative to positive due to the dominance of dark energy over 
non-relativistic matter. 
As $w_{\mathrm{DE}} \approx w_{\mathrm{eff}}$ in the farther future, 
$w_{\mathrm{DE}}$ oscillates around the phantom divide line 
$w_{\mathrm{DE}}=-1$ because the sign of $\dot{H}$ changes 
and consequently multiple crossings 
can be realized. 

Finally, we mention that in our numerical calculations, 
we have taken the initial conditions  of   $z_0=8.0$, $8.0$, $3.0$ 
and $3.5$ for the models of (i)--(iv)
at $z=z_0$, respectively, so that 
$RF^{\prime}(z=z_0) \sim 10^{-13}$ with $F^{\prime}=dF/dR$,
to ensure that they
 can be all close enough to the $\Lambda\mathrm{CDM}$ model
with $RF^{\prime} =0$. 



In this talk, 
we have explored the past and future evolutions of $w_{\mathrm{DE}}$ 
in the viable $f(R)$ gravity models and explicitly shown that 
the crossings of the phantom divide 
are the generic feature in these models. 
We have demonstrated that
 in the future 
the sign of $\dot{H}$ changes from negative to positive due to the dominance 
of dark energy over non-relativistic matter. 
This is a common physical phenomena to the existing viable $f(R)$ 
models and thus it is one of the peculiar properties  of  $f(R)$ gravity 
models characterizing the deviation 
from the $\Lambda\mathrm{CDM}$ model.

\section*{Acknowledgments}

The work is supported in part by 
the National Science Council of R.O.C. under
Grant \#s: NSC-95-2112-M-007-059-MY3 and
NSC-98-2112-M-007-008-MY3
and 
National Tsing Hua University under the Boost Program and Grant \#: 
99N2539E1.


\end{document}